# Davydov Splitting and Excitonic Resonance Effects in Raman Spectra of Few-Layer MoSe$_2$

*Kangwon Kim,*[†,1] *Jae-Ung Lee,*[†,1] *Dahyun Nam,*[†] *and Hyeonsik Cheong*[†,*]

[†]Department of Physics, Sogang University, Seoul 04107, Korea

**Corresponding Author**

*hcheong@sogang.ac.kr

ABSTRACT

Raman spectra of few-layer MoSe$_2$ were measured with 8 excitation energies. New peaks that appear only near resonance with various exciton states are analyzed, and the modes are assigned. The resonance profiles of the Raman peaks reflect the joint density of states for optical transitions, but the symmetry of the exciton wave functions leads to selective enhancement of the $A_{1g}$ mode at the A exciton energy and the shear mode at the C exciton energy. We also find Davydov splitting of *intra*-layer $A_{1g}$, $E_{1g}$, and $A_{2u}$ modes due to *inter*-layer interaction for some excitation energies near resonances. Furthermore, by fitting the spectral positions of *inter*-layer shear and breathing modes and Davydov splitting of *intra*-layer modes to a linear chain model, we extract the strength of the *inter*-layer interaction. We find that the second-nearest-neighbor inter-layer interaction amounts to about 30% of the nearest-neighbor interaction for both in-plane and out-of-plane vibrations.

Few-layer semiconducting transition metal dichalcogenides (TMD's) are studied intensively owing to bandgap energies in the range of near infrared to visible wavelengths which make them suitable for various electronic and optoelectronic applications.[1,2] Monolayer $MoSe_2$ shows a luminescence peak at ~1.6 eV which is suitable for applications in deep red or near infrared regions of the spectrum. $MoSe_2$ is also used in TMD heterostructures such as $WSe_2$/$MoSe_2$[3,4] or $MoS_2$/$MoSe_2$[5,6] which exhibit interesting physical properties due to unique band alignment between these atomically thin semiconductors. Raman spectroscopy is a powerful tool to characterize 2-dimensional materials such as graphene or TMD's. For TMD materials, Raman spectroscopy is used to characterize the number of layers,[7–9] the stacking order,[10–12] strain,[13,14] or doping density.[15] However, it has been reported that the Raman spectrum of a TMD material varies greatly depending on the excitation laser used, which is attributed to excitonic resonance effects.[16–19] Because of reduced dielectric screening in 2-dimensional materials, the excitons in TMD's are known to have very large binding energies,[20–24] and so tightly localized wave functions. Resonance with such exciton states greatly modifies the Raman scattering process to result in extraordinary resonance Raman effects. In the Raman spectra of $MoTe_2$ and $WS_2$, Davydov splitting of some of the main Raman peaks have been reported.[25–27] Davydov splitting, also known as factor-group splitting, is the splitting of bands in the electronic or vibrational spectra of crystals due to the presence of more than one equivalent entity in the unit cell.[28,29] Since this is due to breaking of degeneracy by interactions of each entity, it is an important probe to investigate interactions between each system. In TMD's, the weak *inter*-layer interaction causes splitting of the *intra*-layer vibration modes. Since this splitting is directly related to the inter-layer interaction, the strength of the inter-layer interaction can be estimated by carefully analyzing the splitting. Furthermore, since the number of split peaks and their relative intensities explicitly correlate with the number of layers, Davydov splitting can be used as a fingerprint for the number of layers. As we will present in this paper, Davydov splitting appears in the spectrum only for some excitation wavelengths, which again is related to the excitonic resonances. Therefore, in order to utilize Raman spectroscopy for characterization of TMD's, establishing the excitation energy dependence of the Raman spectrum for each material is an important first step. Here we present Raman spectroscopy on few-layer MoSe$_2$ using 8 different excitation energies and report on resonance effects on the Raman spectra and Davydov splitting of the intra-layer Raman modes. We extract the strength of the *inter*-layer interaction by fitting the spectral positions of *inter*-layer shear and breathing modes and Davydov splitting of *intra*-layer modes.

RESULTS AND DISCUSSION

**Raman Spectra of MoSe$_2$.** The crystal structure of 2H-MoSe$_2$ is schematically shown in Figure 1(a). Although there are two polytypes of MoSe$_2$, the samples presented in this work were all 2H polytypes based on the low-frequency Raman spectrum.[10,11] Single-layer MoSe$_2$ is composed of two Se layers and one Mo layer with hexagonal structures to form a trilayer (TL). As shown in the side view, the adjacent layers are rotated by 180º in the 2H polytype. Each TL is connected to the neighboring TL's by weak van der Waals interactions. The Raman spectra of few-layer MoSe$_2$ measured with the excitation energy of 2.41 eV (514.5 nm) are shown in Figure 1(b) and (c). The peak marked by * at −14 cm$^{-1}$, is due to a plasma line of the laser. In the low-frequency range, breathing and shear modes which correspond to rigid out-of-plane and in-plane vibrations of the TL's, respectively, are observed. The shear modes are indicated by red circles, and the breathing modes by black diamonds. These modes are unambiguous fingerprints of the number of layers and stacking orders of layered materials.[8–12] The measured samples all belong to the 2H polytype,[10,11] and the positions of the low-frequency modes are summarized in Figure 1(d). The main $A_{1g}$ and $E_{2g}^1$ peaks are shown in Figure 1(c), and the peak positions are summarized in Figure 1(e). It should be noted that the notations of the modes depend on the number of TL's,[8,9,25,30] and we use corresponding bulk notations for simplicity unless noted otherwise. The $A_{1g}$ mode blueshifts as the number of TL's increases, whereas the $E_{2g}^1$ mode redshifts with the number of TL's, which is similar to the case of $MoS_2$.[7] This redshift of the $E_{2g}^1$ mode in TMD materials has been attributed to the dielectric screening of the long-range Coulomb interaction and the surface effects.[31,32] For samples thicker than 2TL, additional peaks appear on the low-frequency side of the $A_{1g}$ peak. These peaks are attributed to Davydov splitting[33,34] of the $A_{1g}$ mode. Detailed analyses of these modes will be presented below.

**Mode Assignments for Resonance Raman Spectra.** For further investigation, we repeated the Raman measurements with eight excitation energies as shown in Figure 2. The Raman spectra show dramatic changes as a function of the excitation energy, which suggests a complicated excitonic resonance behavior. The complete set of Raman spectra for all thicknesses are compiled in Supporting Information (Figures S1–7). For 1TL, the Raman peaks are not resolved for 1.58 eV excitation because the spectrum is overwhelmed by the excitonic luminescence. In the low-frequency region, no peaks are observed for 1TL-$MoSe_2$, whereas shear and breathing modes are observed for few-TL samples. The Raman intensities of shear and breathing modes are strongly dependent on the excitation energies. For the excitation energy of 1.58 eV, low-frequency modes are too weak to be resolved except for thicker samples (6–8TL).

For other excitation energies, the shear modes are clearly resolved, but the breathing modes appear strong only for 2.71 and 2.81 eV excitation energies.

In the higher frequency region, several peaks appear or disappear depending on the excitation energy. For example, a peak at ~291 cm$^{-1}$ near the $E_{2g}^1$ mode is observed for excitation energies of 1.58 and 1.96 eV. This peak cannot be due to Davydov splitting because it is observed even in 1TL. This peak was observed before and was interpreted as being due to the splitting of LO and TO phonons of the $E_{2g}^1$ mode.[18,35] The split TO and LO phonons should have polarization dependences similar to those of split $E$ modes under uniaxial strain: the polarization dependence of the two phonons should have anti-correlation.[14,36] In Figure S8, we compare two spectra of 3TL MoSe$_2$ measured with the 1.96 eV excitation in parallel and cross polarizations. The $E_{2g}^1$ mode is clearly observed in both polarizations whereas the peak at ~291 cm$^{-1}$ is observed only for the parallel polarization. This implies that the $E_{2g}^1$ mode peak and the peak at ~291 cm$^{-1}$ are not TO-LO pairs. We note that the frequency of this peak is close to the sum of the frequencies of the TA and $E_{1g}$ branch near the $M$ point of the Brillouin zone.[35] Since 1.58 eV excitation is resonant with the A exciton state, 2-phonon scattering can be enhanced selectively for phonons with specific momenta. Similar resonance effects have also been observed in resonant Raman scattering in MoS$_2$.[16,37–39]

In bulk MoSe$_2$, the $E_{1g}$ mode at ~169 cm$^{-1}$ is forbidden in backscattering, and the $A_{2u}$ mode at ~353 cm$^{-1}$ is Raman inactive; but they appear for some excitation energies due to resonance effects.[40] When the excitation energy is close to the exciton energy, the strongly localized exciton wave function effectively breaks the symmetry of the system, which in turn activates some modes that are normally forbidden. For 1TL-MoSe$_2$, the $E''$ mode corresponding to the bulk $E_{1g}$ mode is also forbidden in back scattering and cannot be observed in the spectrum. The $A_2''$ mode corresponding to the $A_{2u}$ mode is Raman inactive, but a small signal can be observed for the excitation energies of 2.71 and 2.81 eV, probably owing to the resonance effect. It is not clear why the $E''$ mode is completely invisible even at resonant excitations. For 2TL or thicker MoSe$_2$, both the $E_{1g}$ and $A_{2u}$ modes have Davydov splitting, and some of the split modes are Raman active,[25] which may explain why these peaks are relatively strong for thicker MoSe$_2$. A close inspection reveals that the Davydov split peaks indeed can be resolved in some cases (See Supporting Information Figures S9 and S10). The broad peaks labelled '*a*' and '*d*' are strongly enhanced for 1.58 eV excitation. 'Peak *a*' is assigned as a combination of the $E_{2g}^1$ and LA modes at the *M* point, and 'peak *d*' is assigned as the $E_{1g}$+LA or $2B_{2g}$ mode at the *M* point.[40] These second order Raman peaks are enhanced due to resonance with the A exciton state as in the case of MoS$_2$.[16,37–39] The feature labelled '*b*' is observed only at specific excitation energies of 2.41–2.81 eV. In bulk MoSe$_2$, the $B_{1u}$ mode which is the Davydov pair of the $A_{1g}$ mode appears at this position. For few-layer MoSe$_2$, this feature corresponds to the Davydov splitting of the $A_{1g}$ mode and depends sensitively on the number of layers. A detailed analysis of this feature is presented below. The peak labelled '*c*' is observed for all thicknesses, and its intensity more or less correlates with that of the $A_{1g}$ mode. This mode was previously assigned to the second-order $2E_{2g}^2$ mode at the *M* point,[40] but the fact that this mode is observed even for 1TL-MoSe$_2$ precludes this possibility because the $E_{2g}^2$ mode does not exist in 1TL. Another possibility is the combination of TA(*M*) and LA(*M*) modes as suggested in Ref. 18.

**Davydov Splitting.** Now, we focus on the feature '*b*' that is related to the splitting of the $A_{1g}$ vibration mode near 240 cm$^{-1}$. Figure 3(a) shows the details of the spectra in this frequency range for 1–6TL MoSe$_2$ (See Supporting Information for data for 7 and 8TL in Figure S11).

Multiple Lorentzian fits are also shown. The spectra are normalized by the intensity of the highest frequency peak. For 1 and 2TL MoSe$_2$, the spectra can be fitted with a single Lorentzian function except for the one for 2TL measured with the 1.58 eV excitation. 2 peaks are resolved for 3 and 4TL, and 3 peaks for 5 and 6TL. These are direct consequences of Davydov splitting which reflects the inter-layer interaction in layered materials. Figure 3(b) illustrates the out-of-plane vibration modes in 2, 3 and 4TL MoSe$_2$. The vibration modes have the identical configuration within each TL in which the Se atoms vibrate symmetrically with respect to the Mo atom which is stationary. The relative phases between the layers are different. If a vibration mode has an inversion center (for even number of TL's) or a mirror symmetry plane (for odd number of TL's), it is Raman active, and IR active otherwise. In 2TL-MoSe$_2$, there are a Raman active mode ($A_{1g}$) in which the two layers vibrate in phase and an infrared active mode ($A_{2u}$) in which the layers vibrate out of phase. Therefore, normally only one $A_{1g}$ peak is observed for 2TL-MoSe$_2$. The small signal on the lower-frequency side of the $A_{1g}$ peak for the 1.58 eV excitation is ascribed to the out-of-phase $A_{2u}$ mode which becomes partially active due to the resonance effect. For 3TL, the two $A_1'$ modes are Raman active, and the $A_2''$ mode is infrared active. Of the two $A_1'$ modes, all the layers vibrate in phase for the $A_1'(1)$ mode, whereas one layer is out of phase for the $A_1'(2)$ mode. Because the (nonresonant) Raman scattering cross section is proportional to the derivative of the susceptibility tensor with respect to displacement, the contributions of out-of-phase vibrations would tend to cancel each other. Hence the $A_1'(1)$ mode is observed for all excitation energies whereas the $A_1'(2)$ mode appears only for some excitation energies near resonances. Similarly for 4TL, of the two Raman active $A_{1g}$ modes, the all-in-phase $A_{1g}(1)$ mode is always observed at the higher frequency. For 3 and 4TL, the infrared active modes are not resolved even for resonant excitations, probably because they are overwhelmed by relatively stronger Raman active modes with similar frequencies. For any number of TL's, there are $\frac{N+1}{2}$ and $\frac{N}{2}$ Raman active modes for odd number layers and even number layers, respectively.[9,27] Davydov splitting also affects the in-plane $E$ modes.[25] However, the splitting is much smaller in MoSe$_2$. Upon close inspection of the spectra, a small splitting of the $E_{1g}$ mode can be extracted from its asymmetric line shape (See Supporting Information Figure S10). No such splitting is resolved for the $E_{2g}^1$ mode.

The observed peak positions are summarized in Figure 3(c). The peak positions can be estimated using the linear chain model,[25,32] which is schematically described in Figure S12 (See Supporting Information for details of the linear chain model calculations). By considering the second-nearest-neighbor interactions and the surface effects,[25,32] we obtained the spring constants by fitting the experimentally obtained peak positions of the Raman modes (shear, breathing, $E_{2g}^1$, $A_{1g}$, $E_{1g}$, and $A_{2u}$) and the Davydov splitting of the $A_{1g}$, $E_{1g}$, and $A_{2u}$ modes to calculated vibrational frequencies. The in-plane vibrations and the out-of-plane vibrations are fitted separately. The results of the fitting are shown in Figure S13. The thickness dependence of the shear and breathing modes can be modelled by considering only the nearest-neighbor interlayer interaction $\beta$, the interaction between S atoms in adjacent TL's.[8,9] However, the Davydov splitting cannot be fitted without considering the second-nearest-neighbor interaction $\gamma$, the interaction between Mo atoms and S atoms in adjacent TL's. The obtained parameters are summarized in Table 1. The parameters for the in-plane vibrations have larger error bars because of the small Davydov splitting for these modes. The second-nearest-neighbor inter-layer interaction is not negligible as $\gamma$ is about 30% of the nearest-neighbor interaction term $\beta$ for both in-plane and out-of-plane vibrations. The surface effect considers the fact that the S atoms on the surface experience slightly different force constants and needs to be included to explain the redshift of the $E_{2g}^1$ and $A_{2u}$ modes with increasing thickness.[32]

**Intensity Resonance Profiles of Raman Modes.** The Raman intensities as a function of the excitation energy for several peaks are shown in Figure 4. Each peak shows a distinct resonance behavior which implies different exciton-phonon interactions. In 1TL MoSe$_2$, the A exciton state at ~1.6 eV is associated with the direct bandgap at the $K$ points in the momentum space,[41,42] and the B exciton state at ~1.8 eV with the spin-orbit-split-off band. There is also an absorption peak at ~2.6 eV[42] which is often called a C exciton state.[18,24,41] The band-to-band transition (creation of an unbound electron-hole pair) between the valence band maximum and the conduction band minimum starts at 2.2–2.4 eV[43,44] and may not be distinguishable from the C exciton state in the optical spectrum. The resonance profiles of the Raman peaks can be understood in relation to these states.[18] The $A_{1g}/A_1'(1)$ mode shows an enhancement near the resonance with the A exciton state. Since the A exciton state is formed mostly from the $d_{z^2}$ orbital of Mo atoms,[17,45] it is strongly modulated by out-of-plane $A_{1g}/A_1'(1)$ vibrations because the excitons are strongly bound with a large exciton binding energy. There is another resonance near ~2.7 eV for 1TL, which redshifts for thicker layers. This resonance can be associated with the C absorption peak. Peak '$d$', which we assigned to $E_{1g}$+LA or $2B_{2g}$, has a similar resonance behavior (See Figure S14 for the resonance profile of peak '$c$'). This is reasonable because such two-phonon scattering tends to be strongly enhanced only for excitation energies near strongly-bound exciton states.[16] The other main peak, $E_{2g}^1$, has a very different resonance behavior. There is no enhancement near the A exciton state at ~1.6 eV, which can be explained in terms of the atomic orbitals comprising the A exciton state ($d_{z^2}$ orbital of Mo) that does not couple strongly with in-plane vibrations of the $E_{2g}^1$ mode.[17] The enhancement near the C absorption peak for 1TL is not as pronounced as in the case of $A_{1g}$. For thicker layers, the enhancement is very weak. On the other hand, this peak becomes stronger at the highest excitation energy of 3.82 eV. This is the only peak that has the highest intensity at this excitation energy. Further studies are needed to elucidate the origin of this enhancement.

The $E_{1g}$ and $A_{2u}$ modes show similar resonance behaviors. As explained earlier, these modes become allowed for 2TL or thicker layers due to Davydov splitting. The resonance profiles of these peaks are similar to the reported absorption spectrum of $MoSe_2$,[41,42,46] which indicates that the resonance behaviors of these peaks are dominated by the joint density of states for optical transitions. The $A_{1g}/A_1'(2)$ mode due to Davydov splitting has a similar resonance profile as the $E_{1g}$ and $A_{2u}$ modes. [See Figure S14 for the resonance profile of $A_{1g}/A_1'(3)$] The resonance profiles of these peaks have a double-peak shape with the spacing between the two maxima being ~0.3 eV, which is similar to the spin-orbit splitting of the valence band maximum at the $K$ point.[41,42] Based on this observation, we interpret the resonance profiles of $E_{1g}$, $A_{2u}$, and $A_{1g}/A_1'(2)$ modes in the following way: the lower energy resonance peak at ~2.4 eV corresponds to the C exciton energy or the band-to-band transition between the valence band and the conduction band near the $K$ point; and the higher energy resonance peak corresponds to the transition between the spin-orbit split-off band to the conduction band.

Both shear and breathing modes have maximum intensity for the 2.81 eV excitation, which was the highest excitation energy available for low-frequency measurements. There is no data for the 2.54 eV or 3.82 eV excitation as we did not have appropriate filters for those laser lines. The shear mode shows resonance near 2.4 eV which is similar to other peaks such as the $A_{1g}/A_1'(2)$ mode. The breathing modes are almost invisible except for the highest excitation energies, which is similar to the case of $MoS_2$.[12,47] Since inter-layer vibrations are dominated by chalcogen to chalcogen interactions, no enhancement of shear and breathing modes at the A exciton state (~1.6 eV) can be understood. On the other hand, since the orbital of the C exciton states are related to the $p_x$ orbitals of the chalcogen atoms,[17] enhancement of the shear mode at the C exciton state (~2.4 eV) seems reasonable.

CONCLUSIONS

In summary, we analyze resonance Raman spectra of 1–8TL $MoSe_2$ and assign new peaks that appear near resonance with various exciton states. The resonance profiles of the Raman peaks reflect the joint density of states for optical transitions as measured by optical absorption spectra, but in some special cases, the symmetry of the exciton wave functions leads to selective enhancement of the $A_{1g}$ mode at the A exciton energy and the shear mode at the C exciton energy. We find Davydov splitting of *intra*-layer $A_{1g}$, $E_{1g}$, and $A_{2u}$ modes due to *inter*-layer interaction for some excitation energies near resonances. Since the number of split peaks and their relative intensities explicitly correlate with the number of layers, Davydov splitting can be used as a fingerprint for the number of layers. Furthermore, by fitting the spectral positions of *inter*-layer shear and breathing modes and Davydov splitting of *intra*-layer modes to a linear chain model, we extract the strength of the *inter*-layer interaction. We find that the second-nearest-neighbor inter-layer interaction amounts to about 30% of the nearest-neighbor interaction for both in-plane and out-of-plane vibrations.

METHODS

Few-layer $MoSe_2$ samples were prepared by mechanical exfoliation from bulk $MoSe_2$ flakes (HQ Graphene) on Si substrates with a 285-nm $SiO_2$ layer. The number of layers was determined by combination of optical contrast, Raman, and PL measurements. An optical microscope image and typical PL spectra are shown in Figure S15. Micro-Raman measurements were performed with 8 different excitation sources: the 325 and 441.6 nm (3.82 and 2.81 eV) lines of a He-Cd laser; the 457.9, 488, and 514.5 nm (2.71, 2.54, and 2.41 eV, respectively) lines of an Ar ion laser; the 532 nm (2.33 eV) line of a diode-pumped-solid-state laser; the 632.8 nm (1.96 eV) line of a He-Ne laser; and the 784.8 nm (1.58 eV) line of a diode laser. Unwanted wavelengths from the lasers were filtered by reflective volume holographic filters (Ondax and OptiGrate). The laser beam was focused onto the sample by using a 50× objective lens (0.8 N.A.) for all excitation wavelengths except for the 325 nm excitation for which a 40× UV objective lens (0.5 N.A.) was used. The scattered light was collected by the same objective lens (backscattering geometry) and dispersed with a Jobin-Yvon Horiba iHR550 spectrometer (1200 groove/mm for 785 nm and 2400 groove/mm for all the other excitation energies). A liquid-nitrogen-cooled back-illuminated charge-coupled-device detector was used. To access the low-frequency range below 50 $cm^{-1}$, volume holographic filters (Ondax and OptiGrate) were used to reject the Rayleigh-scattered light. The laser power was kept at 50 µW for all measurements to avoid local heating of the samples. The spectral resolution was below 1 $cm^{-1}$. For resonance studies, normalization was carried out to extract the intrinsic resonance effects. First, the Raman intensities are normalized by the intensity of the Si Raman peak at 520 cm$^{-1}$ for each excitation energy to correct for the efficiency of the detection system. Also, the effect of multiple interference from the substrate[48] and the resonance Raman effect of Si[49] are considered. The wavelength dependent refractive indices are taken from the measured values for 1TL MoSe$_2$.[46] Due to the limitations of available data, bulk values[50] were used for the refractive indices in the UV region (for excitation wavelength of 325 nm). The detailed normalization procedure is presented in Ref. 16.

TABLES

**Table 1.** Force constants per unit area obtained by fitting experimental data to linear chain model.

|  | Corresponding interaction | Out-of-plane | In-plane |
|---|---|---|---|
| $\alpha$ ($10^{19}$ N/m³) | Intra-layer Mo-Se | 229±1 | 152±1 |
| $\alpha_e$ ($10^{19}$ N/m³) | Intra-layer Mo-Se for surface Se | 235±1 | 155±1 |
| $\beta$ ($10^{19}$ N/m³) | Inter-layer Se-Se | 5.47±0.2 | 1.82±0.4 |
| $\gamma$ ($10^{19}$ N/m³) | Inter-layer Mo-Se | 1.63±0.05 | 0.55±0.1 |
| $\delta$ ($10^{19}$ N/m³) | Intra-layer Se-Se | 2.53±0.04 | −6.80±0.5 |
| $\delta_e$ ($10^{19}$ N/m³) | Intra-layer Se-Se for surface Se | 2.53±0.04 | −6.70±0.5 |

## ASSOCIATED CONTENT

**Supporting Information Available**: Complete set of Raman spectra for all thicknesses. Polarized Raman spectra of 3TL MoSe$_2$ with the 1.96 eV excitation energy. Additional Raman spectra showing Davydov splitting for $A_{2u}$, $E_{1g}$, and $A_{1g}$ modes. Calculation of vibrational mode frequencies based on linear chain model. Resonance profiles for peak '$c$' and $A_{1g}/A_1'(3)$. Optical microscope image of MoSe$_2$ sample and photoluminescence spectra of few-layer MoSe$_2$. This material is available free of charge *via* the Internet at http://pubs.acs.org.

## AUTHOR INFORMATION

**Corresponding Author**

*E-mail: hcheong@sogang.ac.kr

**Author Contributions.**

K.K. and J.-U.L. carried out experimental measurements and analyzed data. All authors interpreted the spectroscopic data together and contributed to preparation of the manuscript. [1]These authors contributed equally.

**Notes**

The authors declare no competing financial interest.

## ACKNOWLEDGMENT

This work was supported by the National Research Foundation (NRF) grant funded by the Korean government (MSIP) (NRF-2016R1A2B300863) and by a grant (No. 2011-0031630) from the Center for Advanced Soft Electronics under the Global Frontier Research Program of MSIP. We thank HC Lee for helpful discussions.

# Supporting Information

# Davydov Splitting and Excitonic Resonance Effects in Raman Spectra of Few-Layer MoSe$_2$

*Kangwon Kim,*[†,1] *Jae-Ung Lee,*[†,1] *Dahyun Nam,*[†] *and Hyeonsik Cheong*[†]

[†]Department of Physics, Sogang University, Seoul 04107, Korea

**Contents:**

- **Figure S1.** Raman spectra of 4–6TL MoSe$_2$ measured with 8 excitation energies.

- **Figure S2.** Raman spectra of 7–8TL MoSe$_2$ measured with 7 excitation energies.

- **Figure S3.** Low-frequency Raman spectra of MoSe$_2$ measured with 6 excitation energies.

- **Figure S4.** Thickness dependence of Raman spectrum of MoSe$_2$ measured with excitation energies 1.58 eV and 1.96 eV.

- **Figure S5.** Thickness dependence of Raman spectrum of MoSe$_2$ measured with excitation energies 2.33 eV and 2.41 eV.

- **Figure S6.** Thickness dependence of Raman spectrum of MoSe$_2$ measured with excitation energies 2.54 eV and 2.71 eV.

- **Figure S7.** Thickness dependence of Raman spectrum of MoSe$_2$ measured with excitation energies 2.81 eV and 3.82 eV.

- **Figure S8.** Polarized Raman spectra of 3TL MoSe$_2$ with 1.96 eV excitation energy.

- **Figure S9.** Normalized Raman spectra near $A_{2u}$ mode showing Davydov splitting for 4, 7 and 8TL MoSe$_2$ for various excitation energies.

- **Figure S10.** Normalized Raman spectra near $E_{1g}$ mode showing Davydov splitting for 4, 7, and 8TL MoSe$_2$ for various excitation energies.

- **Figure S11.** Normalized Raman spectra near $A_{1g}$ mode showing Davydov splitting for 7 and 8TL MoSe$_2$ for various excitation energies.

- **Calculation of vibrational mode frequencies based on linear chain model**

- **Figure S12.** Schematic of linear chain model.

- **Figure S13.** Peak positions of vibrational modes from experiment and linear chain model.

- **Figure S14.** Raman intensity as a function of excitation energies for peak '*c*' and $A_{1g}/A_1'(3)$ mode.

- **Figure S15.** Optical microscope image of MoSe$_2$ sample and photoluminescence (PL) spectra of few-layer MoSe$_2$.

- **References**

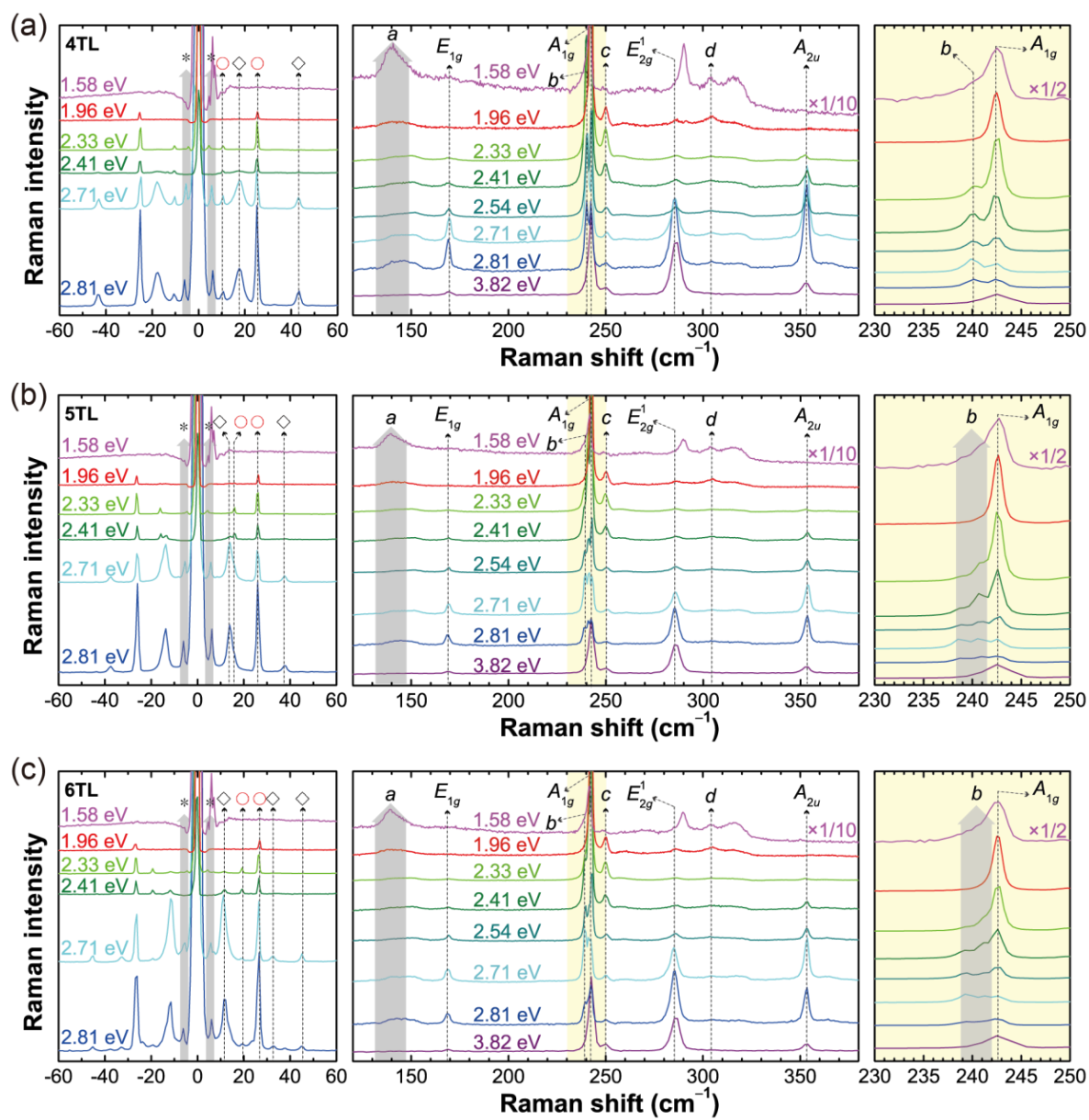

**Figure S1.** Raman spectra of (a) 4TL, (b) 5TL, and (c) 6TL MoSe$_2$ measured with 8 excitation energies indicated. Detailed line shapes of the $A_{1g}$ mode are shown on the right.

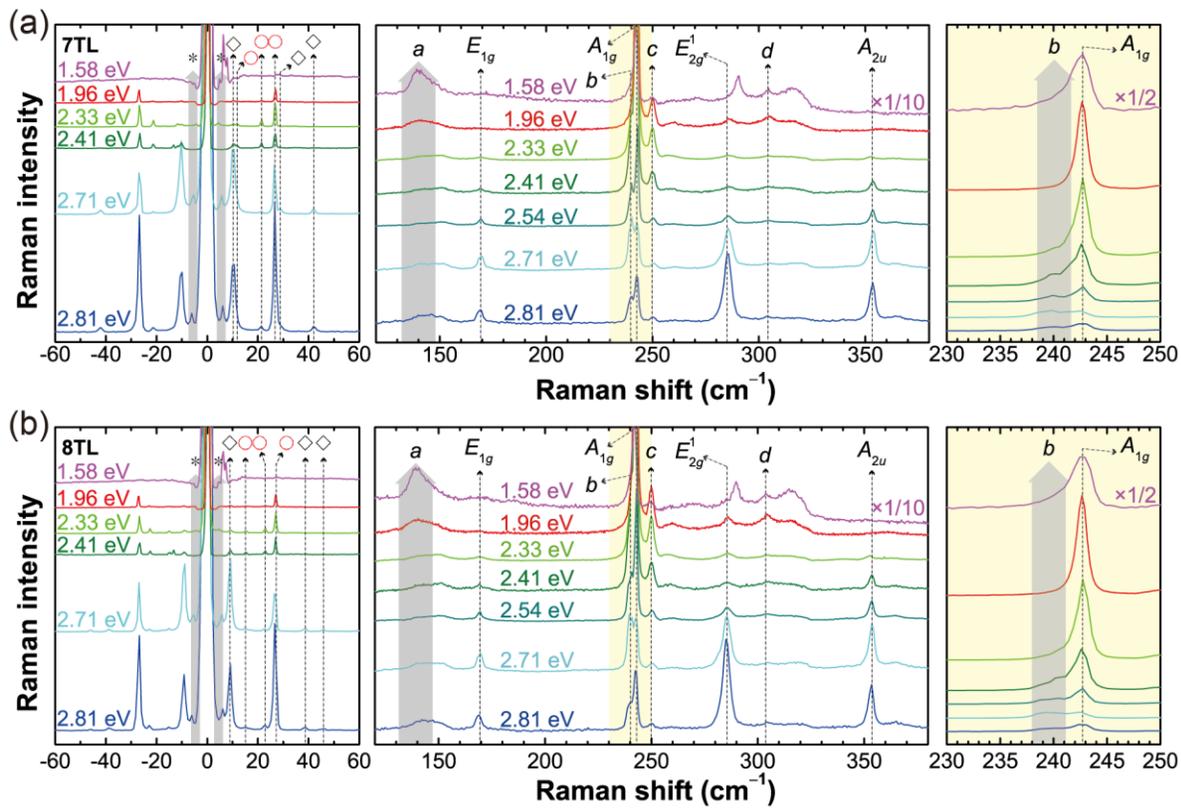

**Figure S2.** Raman spectra of (a) 7TL and (b) 8TL MoSe$_2$ measured with 7 excitation energies indicated. Detailed line shapes of the $A_{1g}$ mode are shown on the right.

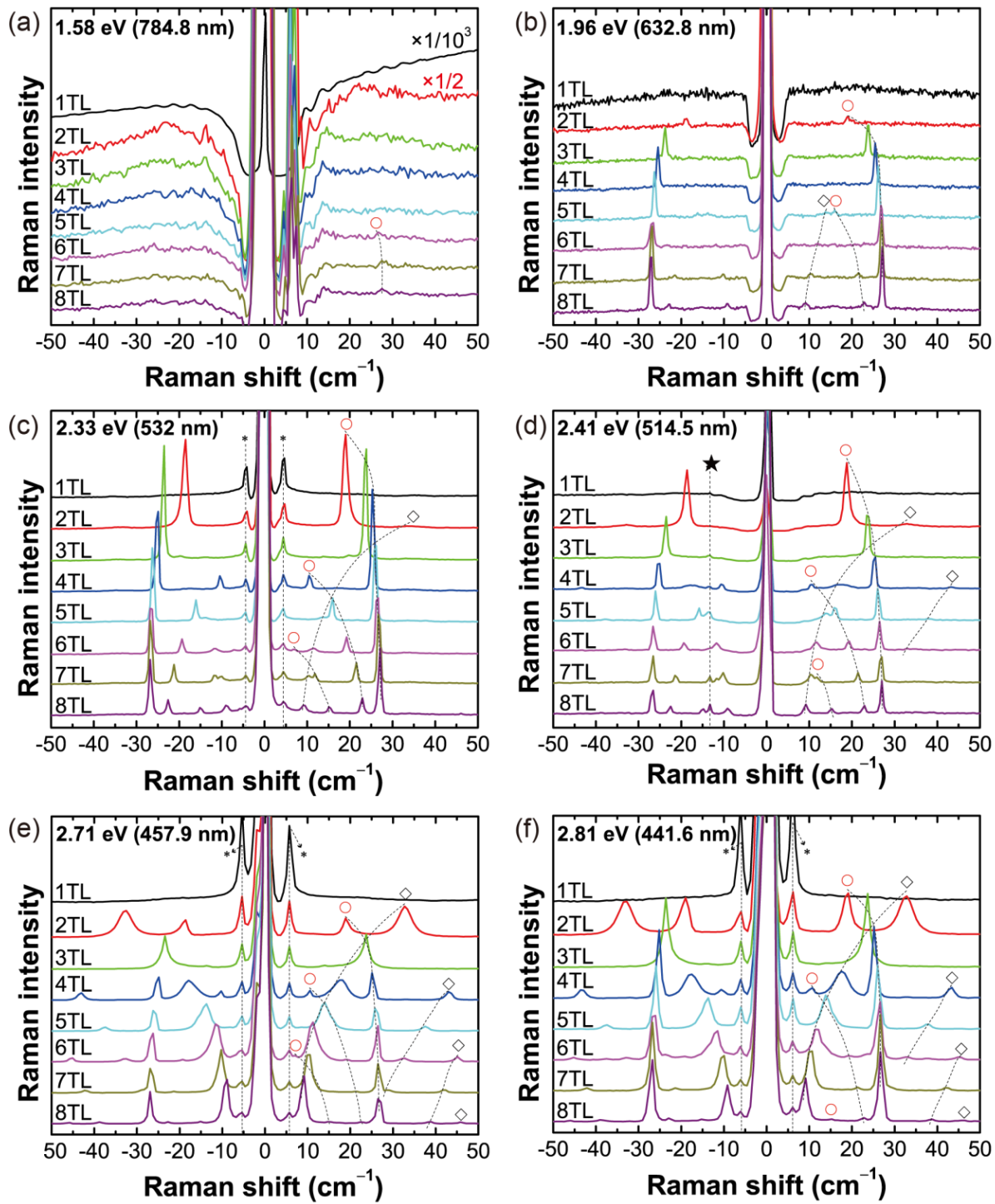

**Figure S3.** Low-frequency Raman spectra of MoSe$_2$ measured with 6 excitation energies indicated. A plasma line of the 2.41 eV laser is indicated by (★), and the Brillouin scattering peak of the Si substrate is indicated by (*).

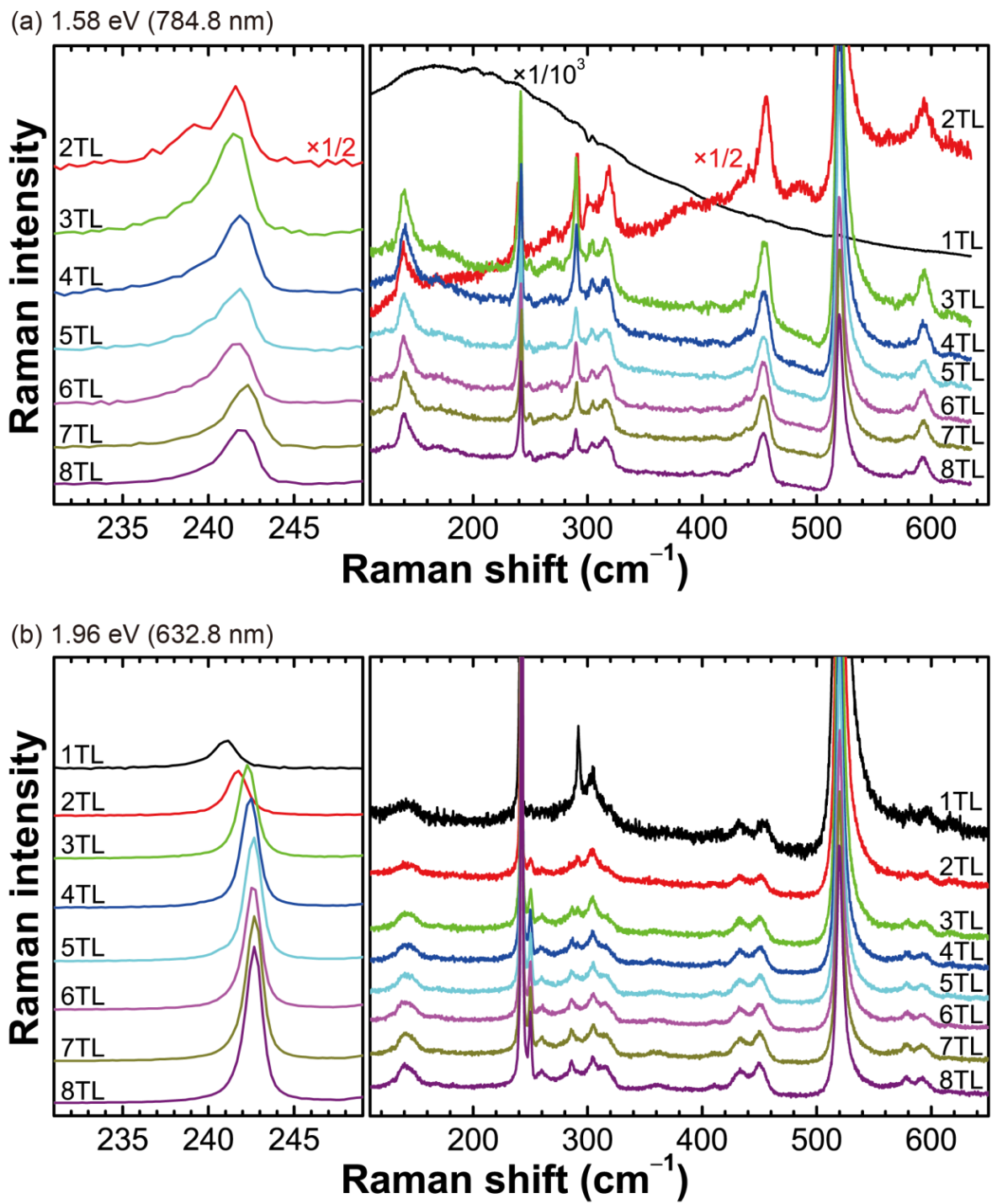

**Figure S4.** Thickness dependence of the Raman spectrum of MoSe$_2$ measured with excitation energies (a) 1.58 eV and (b) 1.96 eV. Detailed line shapes of the $A_{1g}$ mode are shown on the left.

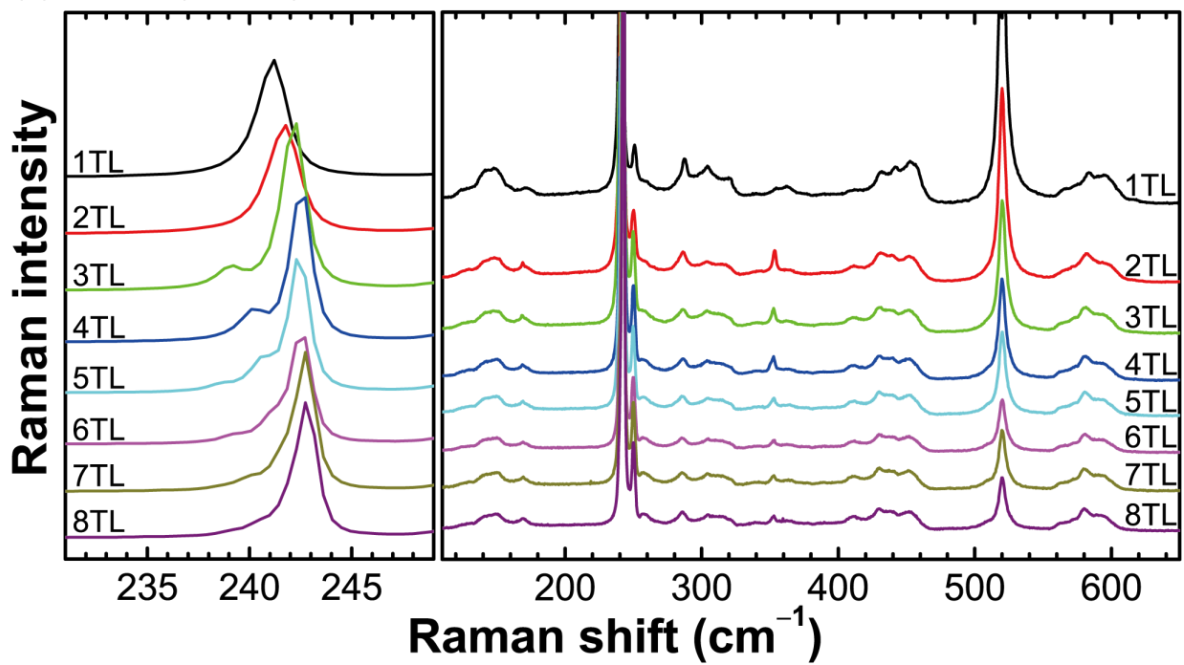
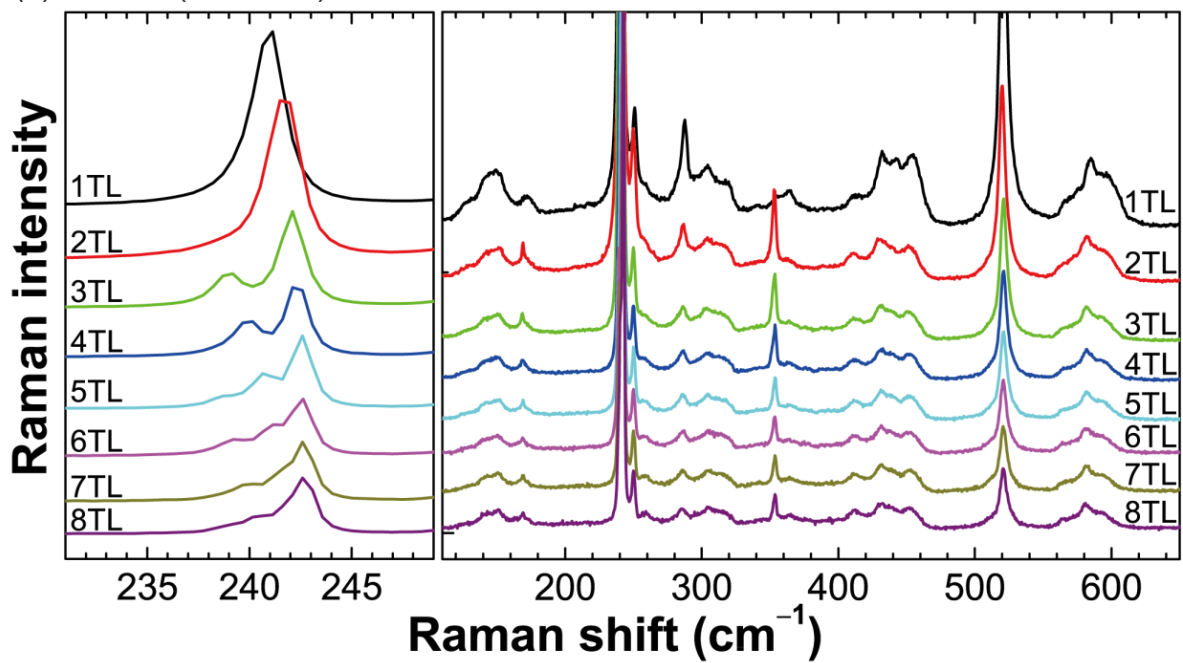

**Figure S5.** Thickness dependence of the Raman spectrum of MoSe$_2$ measured with excitation energies (a) 2.33 eV and (b) 2.41 eV. Detailed line shapes of the $A_{1g}$ mode are shown on the left.

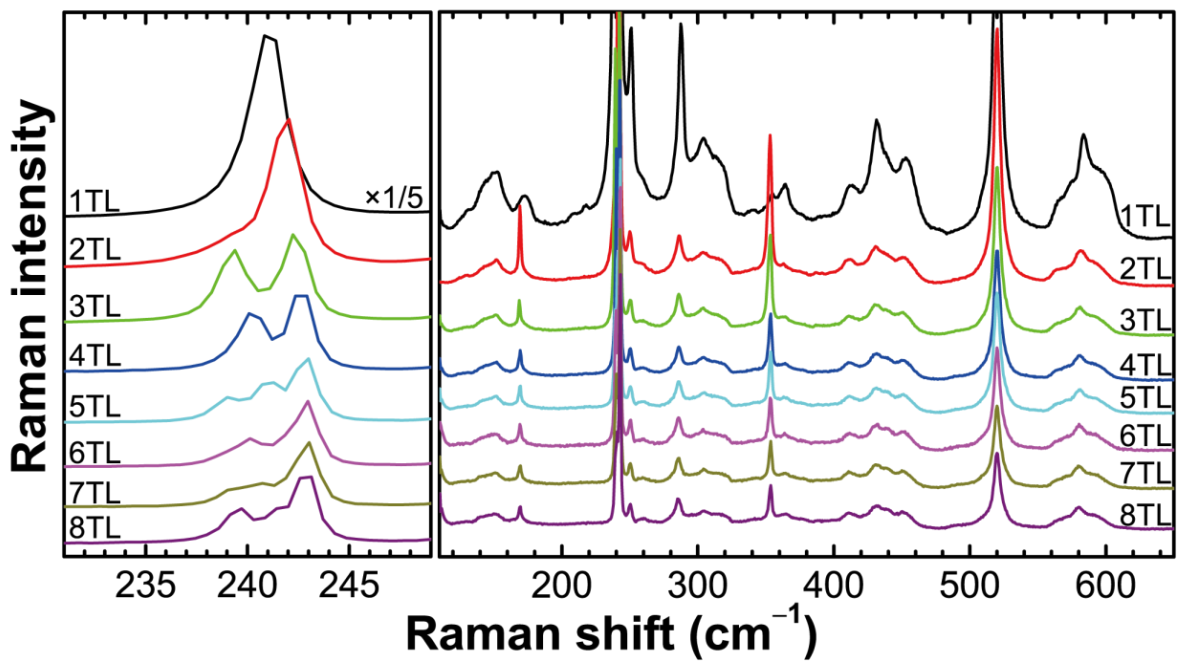
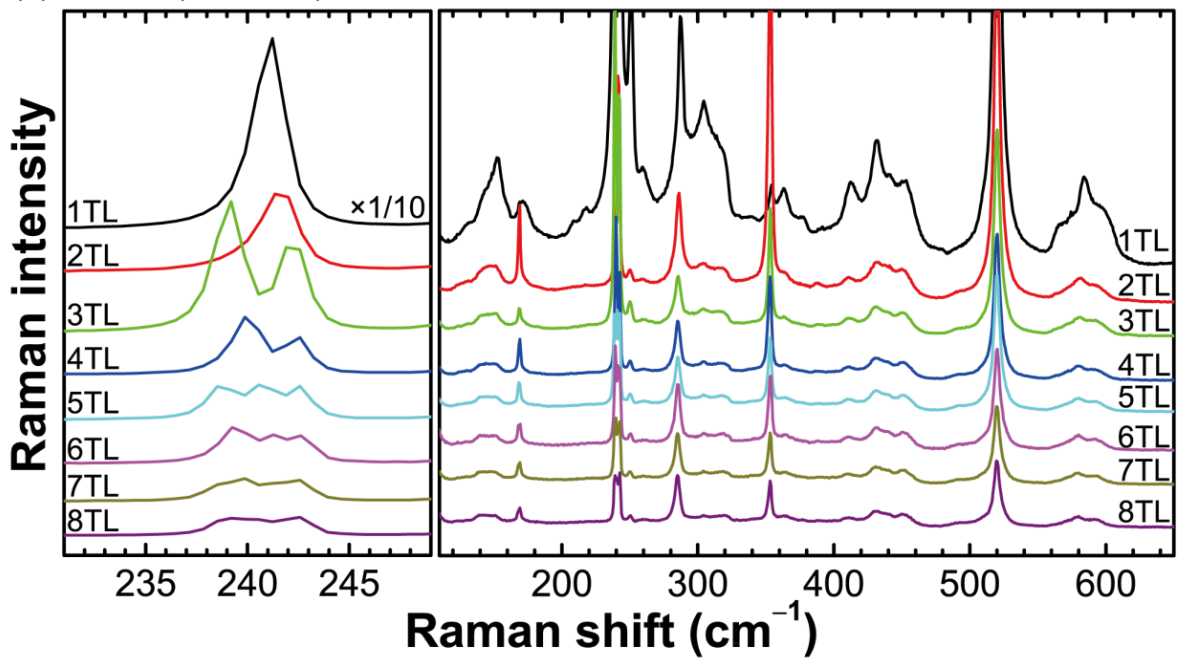

**Figure S6.** Thickness dependence of the Raman spectrum of MoSe$_2$ measured with excitation energies (a) 2.54 eV and (b) 2.71 eV. Detailed line shapes of the $A_{1g}$ mode are shown on the left.

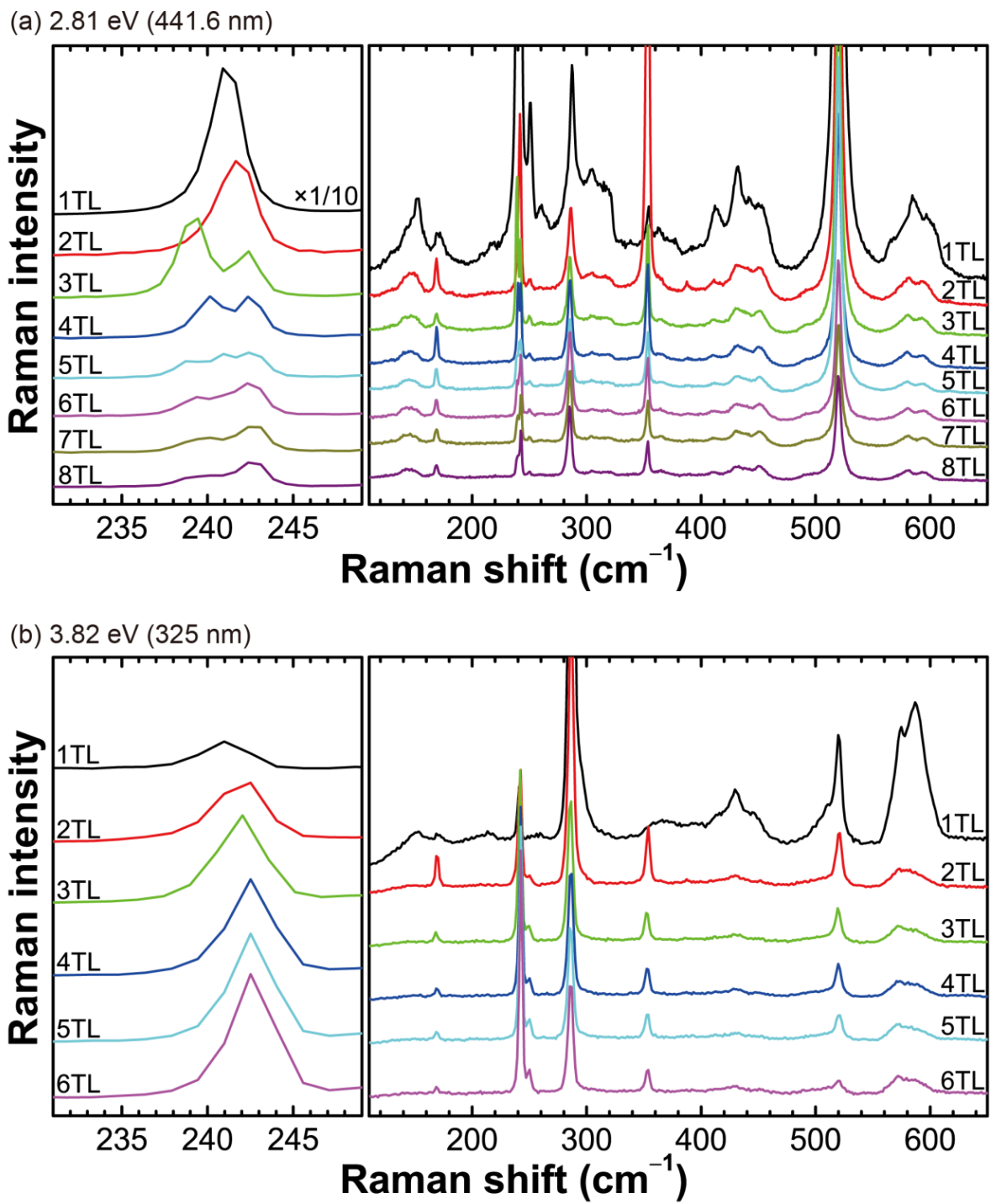

**Figure S7.** Thickness dependence of the Raman spectrum of MoSe$_2$ measured with excitation energies (a) 2.81 eV and (b) 3.82 eV. Detailed line shapes of the $A_{1g}$ mode are shown on the left.

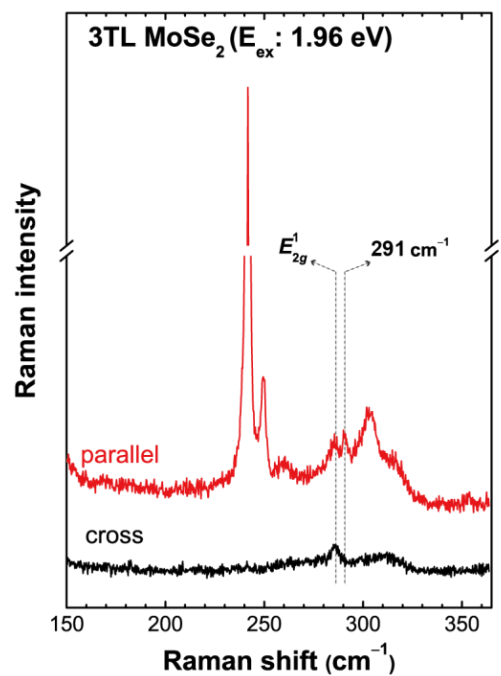

**Figure S8.** Polarized Raman spectra of 3TL MoSe$_2$ with the 1.96 eV excitation energy. The $E_{2g}^1$ mode appears in both polarizations whereas the peak at 291 cm$^{-1}$ disappears in cross polarization.

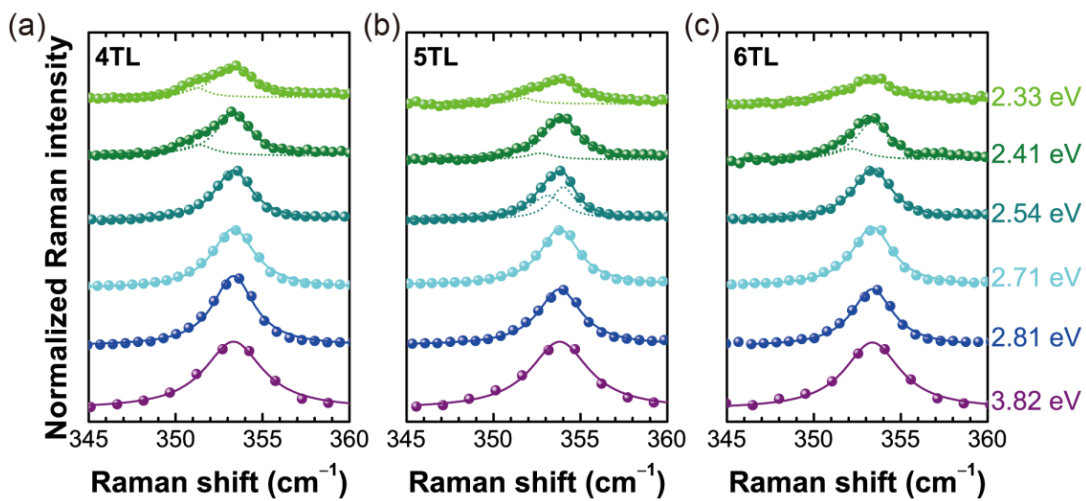

**Figure S9.** Normalized Raman spectra near the $A_{2u}$ mode showing Davydov splitting in (a) 4TL, (b) 5TL, and (c) 6TL $MoSe_2$ for various excitation energies.

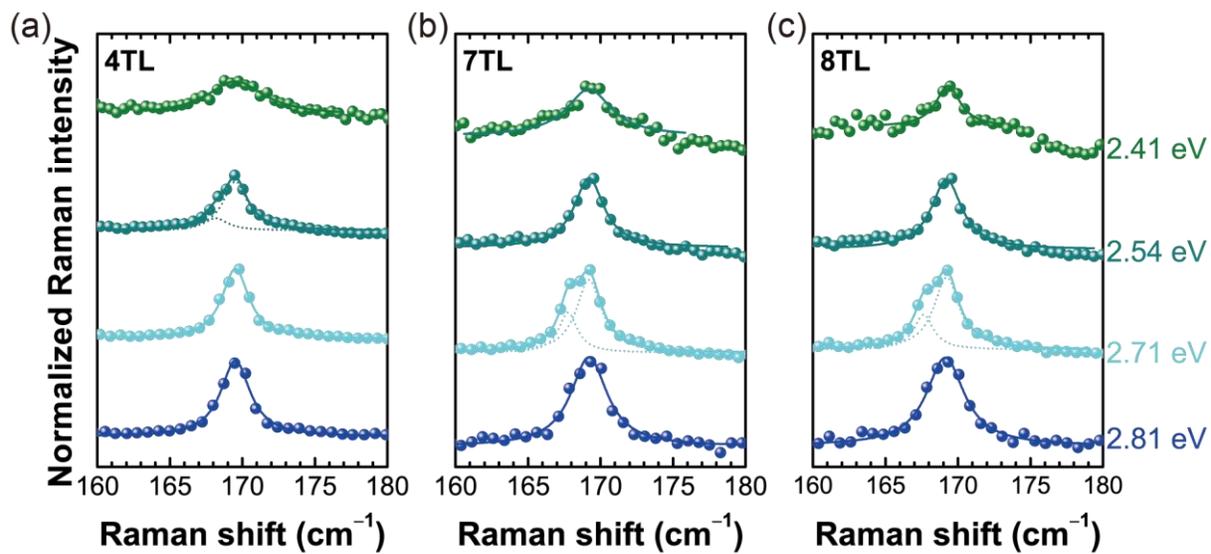

**Figure S10.** Normalized Raman spectra near the $E_{1g}$ mode showing Davydov splitting in (a) 4TL, (b) 7TL, and (c) 8TL MoSe$_2$ for various excitation energies. The splitting could not be resolved for 5TL or 6TL.

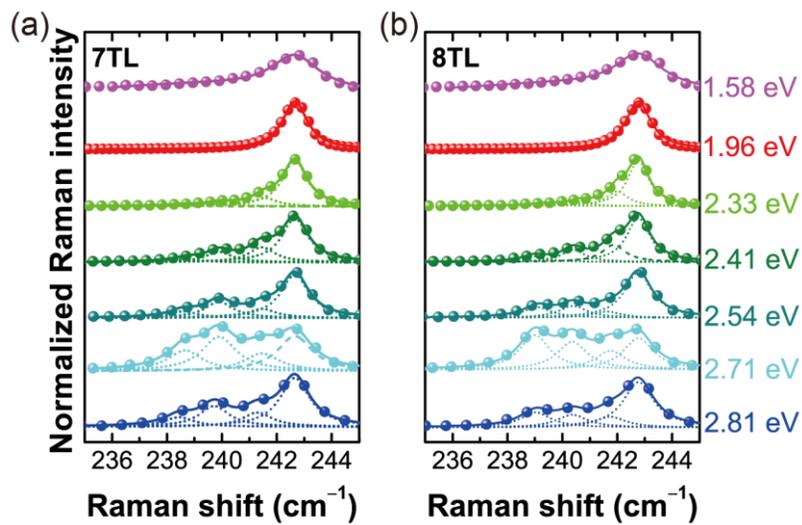

**Figure S11.** Normalized Raman spectra near the $A_{1g}$ mode showing Davydov splitting in (a) 7TL and (b) 8TL MoSe$_2$ for various excitation energies.

**Calculation of vibrational mode frequencies based on linear chain model**

The vibrational modes can be described by the linear chain model.[1,2] We assume that the elastic response of the crystal is a linear function of the forces. By assuming that each atom is connected with other atoms by springs, one can formulate the equations of motion that describe the vibrational modes. Up to second-nearest-neighbor interactions and surface effects are considered. The schematic of the model for the case of 3TL is shown in Figure S12. The vibrational modes along the in-plane and out-of-plane directions are considered separately.

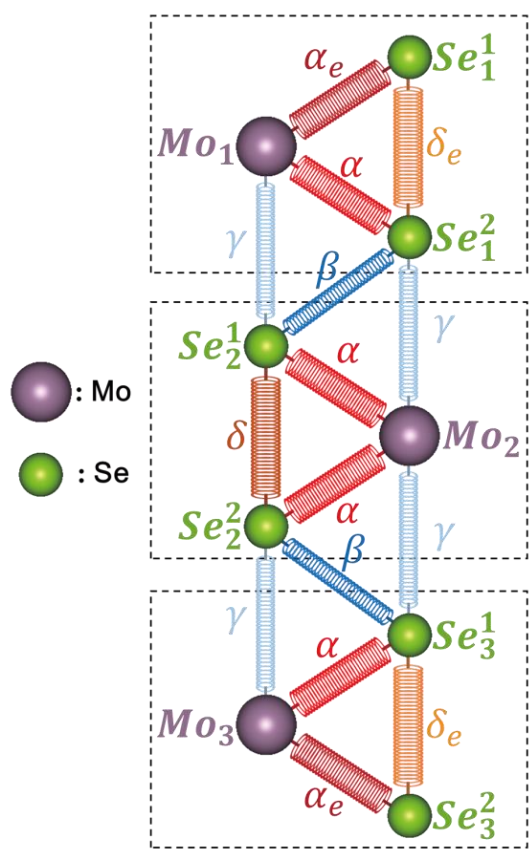

**Figure S12.** Schematic of linear chain model.

The equations of motion can be derived by the following procedure. For example, a Mo atom in the first layer (Mo$_1$) is connected by three springs. The force applied to this Mo atom can be described by

$$F_{Mo_1} = \mu_{Mo}\frac{d^2 u_{Mo_1}}{dt^2} = \alpha_e(u_{Mo_1} - u_{Se_1^1}) + \alpha(u_{Mo_1} - u_{Se_1^2}) + \gamma(u_{Mo_1} - u_{Se_2^1})$$

Similarly, the force applied to Se atoms in the first layer can be written as follows.

$$F_{Se_1^1} = \mu_{Se}\frac{d^2 u_{Se_1^1}}{dt^2} = \alpha_e(u_{Se_1^1} - u_{Mo_1}) + \delta_e(u_{Se_1^1} - u_{Se_1^2})$$

$$F_{Se_1^2} = \mu_{Se}\frac{d^2 u_{Se_1^2}}{dt^2} = \delta_e(u_{Se_1^2} - u_{Se_1^1}) + \alpha(u_{Se_1^2} - u_{Mo_1}) + \beta(u_{Se_1^2} - u_{Se_2^1}) + \gamma(u_{Se_1^2} - u_{Mo_2})$$

By considering the forces applied to atoms in each layer, a 9×9 matrix can be formulated for 3TL-MoSe$_2$. The dynamical matrix $D$ can be written as follows.

$$D = \begin{pmatrix}
\dfrac{\alpha_e+\delta_e}{\mu_{Se}} & -\dfrac{\alpha_e}{\mu_{Se}} & -\dfrac{\delta_e}{\mu_{Se}} & 0 & 0 & 0 & 0 & 0 & 0 \\
-\dfrac{\alpha_e}{\mu_{Mo}} & \dfrac{\alpha_e+\alpha+\gamma}{\mu_{Mo}} & -\dfrac{\alpha}{\mu_{Mo}} & -\dfrac{\gamma}{\mu_{Mo}} & 0 & 0 & 0 & 0 & 0 \\
-\dfrac{\delta_e}{\mu_{Se}} & -\dfrac{\alpha}{\mu_{Se}} & \dfrac{\alpha+\beta+\gamma+\delta_e}{\mu_{Se}} & -\dfrac{\beta}{\mu_{Mo}} & -\dfrac{\gamma}{\mu_{Se}} & 0 & 0 & 0 & 0 \\
0 & -\dfrac{\gamma}{\mu_{Se}} & -\dfrac{\beta}{\mu_{Mo}} & \dfrac{\alpha+\beta+\gamma+\delta}{\mu_{Se}} & -\dfrac{\alpha}{\mu_{Se}} & -\dfrac{\delta}{\mu_{Se}} & 0 & 0 & 0 \\
0 & 0 & -\dfrac{\gamma}{\mu_{Mo}} & -\dfrac{\alpha}{\mu_{Mo}} & 2\dfrac{\alpha+\gamma}{\mu_{Mo}} & -\dfrac{\alpha}{\mu_{Mo}} & -\dfrac{\gamma}{\mu_{Mo}} & 0 & 0 \\
0 & 0 & 0 & -\dfrac{\delta}{\mu_{Se}} & -\dfrac{\alpha}{\mu_{Se}} & \dfrac{\alpha+\beta+\gamma+\delta}{\mu_{Se}} & -\dfrac{\beta}{\mu_{Mo}} & -\dfrac{\gamma}{\mu_{Se}} & 0 \\
0 & 0 & 0 & 0 & -\dfrac{\gamma}{\mu_{Se}} & -\dfrac{\beta}{\mu_{Mo}} & \dfrac{\alpha+\beta+\gamma+\delta_e}{\mu_{Se}} & -\dfrac{\alpha}{\mu_{Mo}} & -\dfrac{\delta_e}{\mu_{Se}} \\
0 & 0 & 0 & 0 & 0 & -\dfrac{\gamma}{\mu_{Mo}} & -\dfrac{\alpha}{\mu_{Se}} & \dfrac{\alpha_e+\alpha+\gamma}{\mu_{Mo}} & -\dfrac{\alpha_e}{\mu_{Se}} \\
0 & 0 & 0 & 0 & 0 & 0 & -\dfrac{\delta_e}{\mu_{Se}} & -\dfrac{\alpha_e}{\mu_{Mo}} & \dfrac{\alpha_e+\delta_e}{\mu_{Se}}
\end{pmatrix}$$

For the normal modes, the equations of motion can be expresses as

$$\frac{d^2 U}{dt^2} = -DU = -\omega^2 U,$$

where $U = (u_{Se_1^1}, u_{Mo_1}, u_{Se_1^2}, u_{Se_2^1}, u_{Mo_2}, u_{Se_2^2}, u_{Se_3^1}, u_{Mo_3}, u_{Se_3^2})$. The frequencies of vibrational modes can be obtained from the eigenvalues of the dynamical matrix $D$. By comparing with the peak frequencies obtained from Raman measurements, the spring constants for in-plane and out-of-plane direction can be obtained separately. The fitted results and the experimental data are compared in Figure S13.

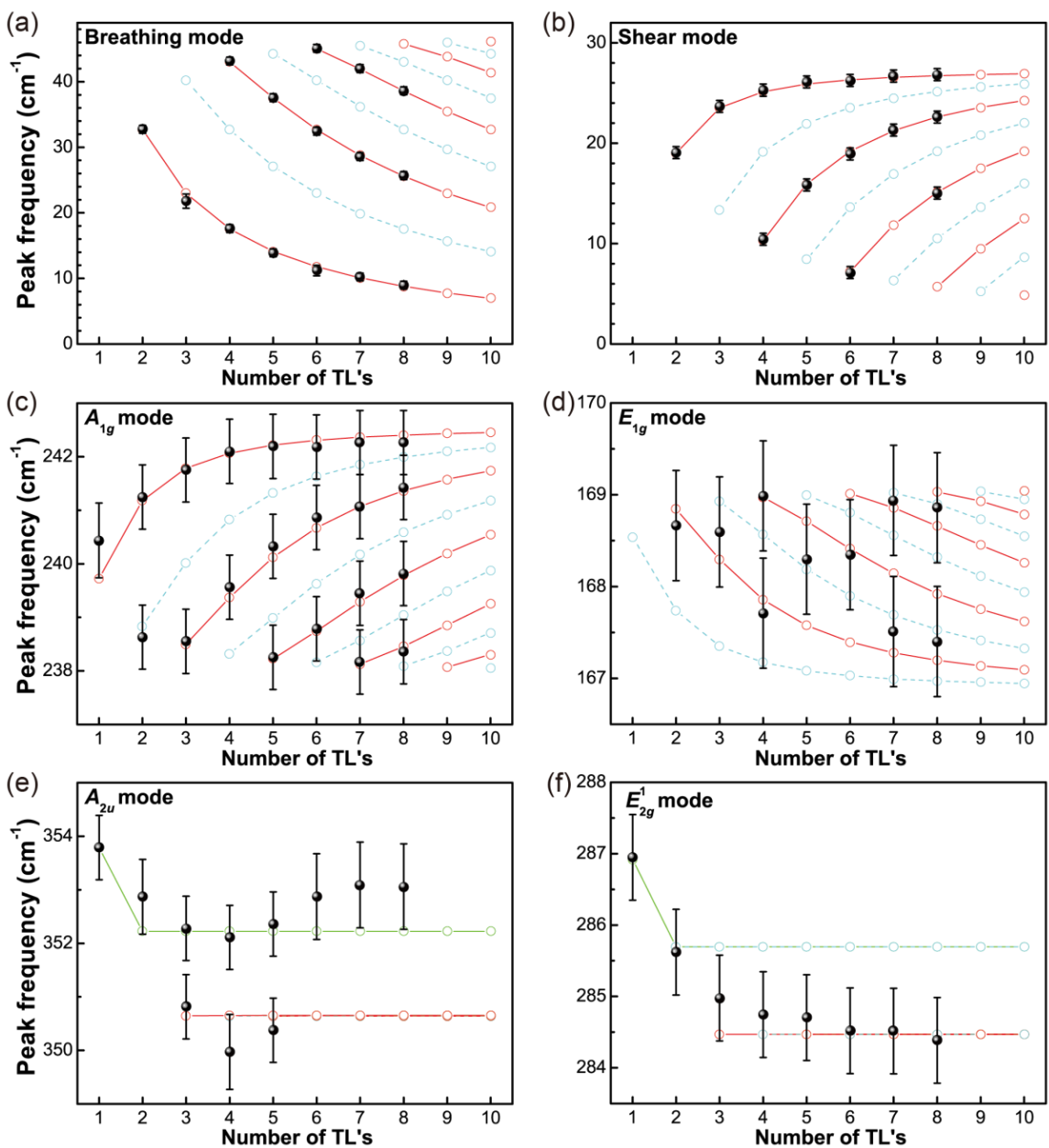

**Figure S13.** Peak positions of vibrational modes from experiment and linear chain model: (a) breathing, (b) shear, (c) $A_{1g}$, (d) $E_{1g}$, (e) $A_{2u}$, and (f) $E_{2g}^1$ modes. Red symbols indicate Raman active modes whereas cyan symbols correspond to either infrared active modes or Raman active modes that are forbidden in backscattering. In (e) and (f), the green data correspond to the Raman active surface mode. The Davydov splitting is too small to be resolved for $A_{2u}$ and $E_{2g}^1$.

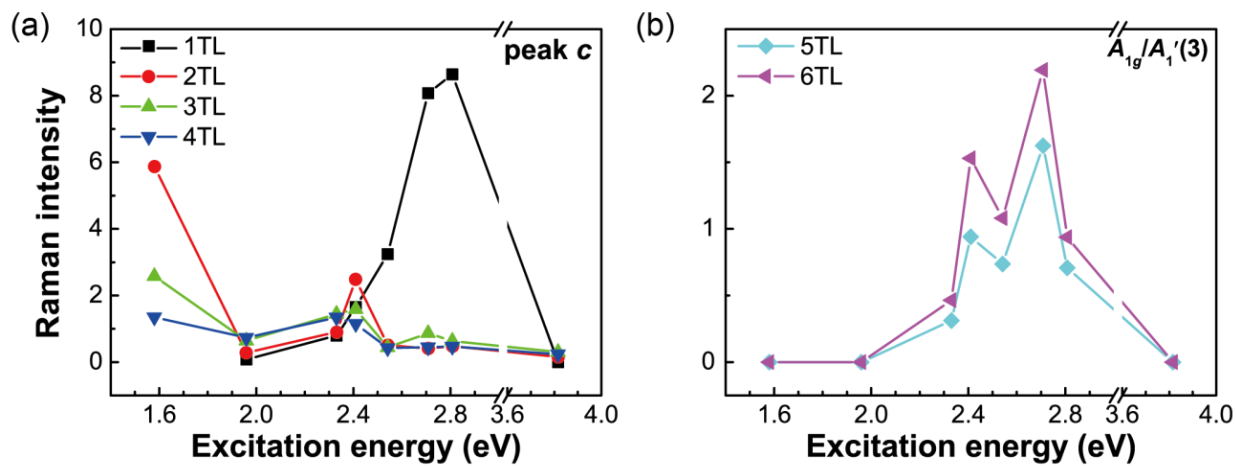

**Figure S14.** Raman intensity as a function of excitation energies for (a) peak '*c*' and (b) $A_{1g}/A_1'(3)$ mode.

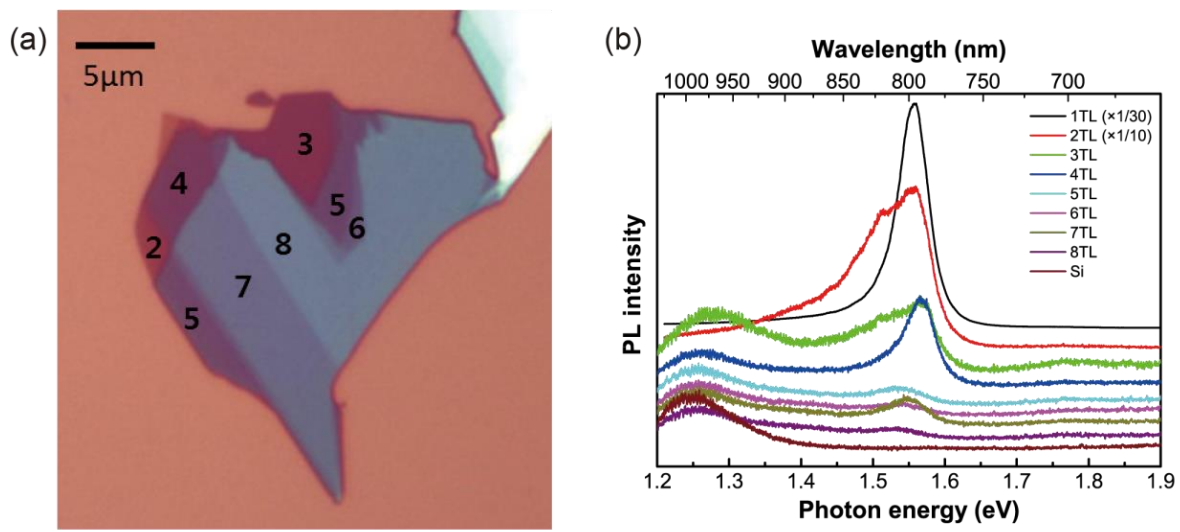

**Figure S15.** (a) Optical microscope image of MoSe$_2$ sample. (b) Photoluminescence (PL) spectra of few-layer MoSe$_2$.